\makeatletter
\@ifundefined{@parse@version@dash}{%
\def\@parse@version#1{\@parse@version@0#1}
\def\@parse@version@#1/#2/#3#4#5\@nil{%
\@parse@version@dash#1-#2-#3#4\@nil}
\def\@parse@version@dash#1-#2-#3#4#5\@nil{%
  \if\relax#2\relax\else#1\fi#2#3#4 }
}{}
\makeatother

\documentclass[%
 reprint,
 amsmath,amssymb,
 aps,
 prl,
]{revtex4-1}
\usepackage{graphicx}
\usepackage{graphicx}
\usepackage{dcolumn}
\usepackage{bm}
\usepackage{lineno}
\usepackage{soul}
\usepackage[normalem]{ulem}

\usepackage{color}
\newcommand{\HS}{^3_{\Lambda}{\rm H}}
\newcommand{\HF}{^4_{\Lambda}{\rm H}}
\newcommand{\HL}{\Lambda \rightarrow p + \pi^-}
\newcommand{\HSTr}{^3_{\Lambda}{\rm H} \rightarrow d + p + \pi^-}
\newcommand{\HSTW}{^3_{\Lambda}{\rm H} \rightarrow {\rm ^3He} + \pi^-}
\newcommand{\HFTW}{^4_{\Lambda}{\rm H} \rightarrow {\rm ^4He} + \pi^-}
\def\snn3{$\sqrt{s_{\rm NN}}$ = 3 GeV}

\usepackage{color}

\usepackage{multirow}

\begin{document}

\title{Observation of Directed Flow of Hypernuclei $^3_{\Lambda}$H and $^4_{\Lambda}$H in \snn3 Au+Au Collisions at RHIC} 
\affiliation{Abilene Christian University, Abilene, Texas   79699}
\affiliation{Alikhanov Institute for Theoretical and Experimental Physics NRC "Kurchatov Institute", Moscow 117218}
\affiliation{Argonne National Laboratory, Argonne, Illinois 60439}
\affiliation{American University of Cairo, New Cairo 11835, New Cairo, Egypt}
\affiliation{Ball State University, Muncie, Indiana, 47306}
\affiliation{Brookhaven National Laboratory, Upton, New York 11973}
\affiliation{University of Calabria \& INFN-Cosenza, Italy}
\affiliation{University of California, Berkeley, California 94720}
\affiliation{University of California, Davis, California 95616}
\affiliation{University of California, Los Angeles, California 90095}
\affiliation{University of California, Riverside, California 92521}
\affiliation{Central China Normal University, Wuhan, Hubei 430079 }
\affiliation{University of Illinois at Chicago, Chicago, Illinois 60607}
\affiliation{Creighton University, Omaha, Nebraska 68178}
\affiliation{Czech Technical University in Prague, FNSPE, Prague 115 19, Czech Republic}
\affiliation{ELTE E\"otv\"os Lor\'and University, Budapest, Hungary H-1117}
\affiliation{Frankfurt Institute for Advanced Studies FIAS, Frankfurt 60438, Germany}
\affiliation{Fudan University, Shanghai, 200433 }
\affiliation{University of Heidelberg, Heidelberg 69120, Germany }
\affiliation{University of Houston, Houston, Texas 77204}
\affiliation{Huzhou University, Huzhou, Zhejiang  313000}
\affiliation{Indian Institute of Science Education and Research (IISER), Berhampur 760010 , India}
\affiliation{Indian Institute of Science Education and Research (IISER) Tirupati, Tirupati 517507, India}
\affiliation{Indian Institute Technology, Patna, Bihar 801106, India}
\affiliation{Indiana University, Bloomington, Indiana 47408}
\affiliation{Institute of Modern Physics, Chinese Academy of Sciences, Lanzhou, Gansu 730000 }
\affiliation{University of Jammu, Jammu 180001, India}
\affiliation{Joint Institute for Nuclear Research, Dubna 141 980}
\affiliation{Kent State University, Kent, Ohio 44242}
\affiliation{University of Kentucky, Lexington, Kentucky 40506-0055}
\affiliation{Lawrence Berkeley National Laboratory, Berkeley, California 94720}
\affiliation{Lehigh University, Bethlehem, Pennsylvania 18015}
\affiliation{Max-Planck-Institut f\"ur Physik, Munich 80805, Germany}
\affiliation{Michigan State University, East Lansing, Michigan 48824}
\affiliation{National Research Nuclear University MEPhI, Moscow 115409}
\affiliation{National Institute of Science Education and Research, HBNI, Jatni 752050, India}
\affiliation{National Cheng Kung University, Tainan 70101 }
\affiliation{Ohio State University, Columbus, Ohio 43210}
\affiliation{Panjab University, Chandigarh 160014, India}
\affiliation{NRC "Kurchatov Institute", Institute of High Energy Physics, Protvino 142281}
\affiliation{Purdue University, West Lafayette, Indiana 47907}
\affiliation{Rice University, Houston, Texas 77251}
\affiliation{Rutgers University, Piscataway, New Jersey 08854}
\affiliation{University of Science and Technology of China, Hefei, Anhui 230026}
\affiliation{South China Normal University, Guangzhou, Guangdong 510631}
\affiliation{Shandong University, Qingdao, Shandong 266237}
\affiliation{Shanghai Institute of Applied Physics, Chinese Academy of Sciences, Shanghai 201800}
\affiliation{Southern Connecticut State University, New Haven, Connecticut 06515}
\affiliation{State University of New York, Stony Brook, New York 11794}
\affiliation{Instituto de Alta Investigaci\'on, Universidad de Tarapac\'a, Arica 1000000, Chile}
\affiliation{Temple University, Philadelphia, Pennsylvania 19122}
\affiliation{Texas A\&M University, College Station, Texas 77843}
\affiliation{University of Texas, Austin, Texas 78712}
\affiliation{Tsinghua University, Beijing 100084}
\affiliation{University of Tsukuba, Tsukuba, Ibaraki 305-8571, Japan}
\affiliation{Valparaiso University, Valparaiso, Indiana 46383}
\affiliation{Variable Energy Cyclotron Centre, Kolkata 700064, India}
\affiliation{Wayne State University, Detroit, Michigan 48201}
\affiliation{Yale University, New Haven, Connecticut 06520}

\author{B.~E.~Aboona}\affiliation{Texas A\&M University, College Station, Texas 77843}
\author{J.~Adam}\affiliation{Czech Technical University in Prague, FNSPE, Prague 115 19, Czech Republic}
\author{J.~R.~Adams}\affiliation{Ohio State University, Columbus, Ohio 43210}
\author{G.~Agakishiev}\affiliation{Joint Institute for Nuclear Research, Dubna 141 980}
\author{I.~Aggarwal}\affiliation{Panjab University, Chandigarh 160014, India}
\author{M.~M.~Aggarwal}\affiliation{Panjab University, Chandigarh 160014, India}
\author{Z.~Ahammed}\affiliation{Variable Energy Cyclotron Centre, Kolkata 700064, India}
\author{A.~Aitbaev}\affiliation{Joint Institute for Nuclear Research, Dubna 141 980}
\author{I.~Alekseev}\affiliation{Alikhanov Institute for Theoretical and Experimental Physics NRC "Kurchatov Institute", Moscow 117218}\affiliation{National Research Nuclear University MEPhI, Moscow 115409}
\author{D.~M.~Anderson}\affiliation{Texas A\&M University, College Station, Texas 77843}
% ...

\collaboration{STAR Collaboration}\noaffiliation

\begin{abstract}
We report here the first observation of directed flow ($v_1$) of the hypernuclei $\HS$ and $\HF$ in midcentral Au+Au collisions at \snn3 at RHIC.
These data are taken as part of the beam energy scan program carried out by the STAR experiment. 
From 165 $\times$ 10$^{6}$ events in 5\%-40\% centrality, about 8400 $\HS$ and 5200 $\HF$ candidates are reconstructed through two- and three-body decay channels.
We observe that these hypernuclei exhibit significant directed flow. 
Comparing to that of light nuclei, it is found that the midrapidity $v_1$ slopes of $\HS$ and $\HF$ follow baryon number scaling, implying that the coalescence is the dominant mechanism for these hypernuclei production in the 3 GeV Au+Au collisions.
\end{abstract}

\maketitle

When a nucleon is replaced by a hyperon ($e.g.$, $\Lambda$, $\Sigma$) with strangeness $S$ = -1, a nucleus is transformed into a hypernucleus which allows for the study of the hyperon-nucleon ($Y$-$N$) interaction. It is well known that two-body $Y$-$N$ and three-body $Y$-$N$-$N$ interactions, especially at high baryon density, are essential for understanding the inner structure of compact stars~\cite{Gerstung:2020ktv,Lonardoni:2014bwa}. New results on precision measurements of $\Lambda$-$p$ elastic scattering from Jefferson Lab~\cite{CLAS:2021gur} and $\Sigma^{-}$-$p$ elastic scattering from J-PARC~\cite{J-PARCE40:2021qxa,J-PARCE40:2021bgw} became available recently, which may help to constrain the equation of state of high density matter inside a neutron star. Until recently, almost all hypernuclei measurements have been carried out with light particle ($e.g.$, $e$, $\pi^{+}$, $K^{-}$) induced reactions~\cite{Gal:2016boi,Hashimoto:2006aw,HKS:2014nmp}, where the $Y$-$N$ interaction around the saturation density is analyzed from spectroscopic properties of hypernuclei.

Utilizing hypernuclei production in heavy-ion collisions to study the $Y$-$N$ interaction and the properties of QCD matter has been a subject of interest in the past decades~\cite{STAR:2010gyg,ALICE:2019vlx,STAR:2017gxa,STAR:2021orx,Saito:2021gao}. However, due to limited statistics, measurements have been mainly focused on the light hypernuclei lifetime, binding energy and production yields~\cite{Chen:2018tnh,STAR:2021orx,STAR:2019wjm}. Thermal model~\cite{Andronic:2010qu} and hadronic transport model with coalescence afterburner~\cite{Steinheimer:2012tb,Aichelin:2019tnk} calculations have predicted abundant production of light hypernuclei in high-energy nuclear collisions, especially at high baryon density. 
Anisotropic flow has been commonly used for studying the properties of matter created in high energy nuclear collisions. 
Because of its genuine sensitivity to early collision dynamics~\cite{Hung:1994eq,Brachmann:1999xt,Steinheimer:2014pfa,Nara:2016phs}, the first order coefficient of the Fourier expansion of the azimuthal distribution in the momentum space, $v_1$, also called the directed flow, has been analyzed for many particles species ranging from $\pi$ mesons to light nuclei~\cite{STAR:2014clz,STAR:2017okv,STAR:2020hya,STAR:2013ayu,STAR:2020dav,Bzdak:2019pkr}.
Collective flow is driven by pressure gradients created in such collisions. Hence, measurements of hypernuclei collectivity make it possible to study the $Y$-$N$ interactions in the QCD equation of state at high baryon density.
 
In this Letter, we report the first observation of directed flow, $v_1$, of $\HS$ and $\HF$ in center-of-mass energy \snn3 Au+Au collisions. The data were collected by the STAR experiment at RHIC with the fixed-target (FXT) setup in 2018. A gold beam of energy 3.85 GeV/u is bombarded on a gold target of thickness 1\% interaction length, located at the entrance of STAR's time-projection chamber (TPC)~\cite{Anderson:2003ur}. The TPC, which is the main tracking detector in STAR, is 4.2 m long and 4 m in diameter, positioned inside a 0.5\,T solenoidal magnetic field along the beam direction. The collision vertex position of each event along the beam direction, $V_{z}$, is required to be within $\pm$2 cm of the target position. An additional requirement on the collision vertex position to be within a radius $r$ of less than 2 cm is imposed to eliminate background events from interactions with the beam pipe. Beam-beam counters (BBC)~\cite{Whitten:2008zz} and the time-of-flight (TOF) detector~\cite{Llope:2012zz} are used to obtain the minimum bias (MB) trigger condition. After event selection, a total of 2.6$\times$10$^8$ MB events are used for further analysis.   

The centrality is determined using the charged particle multiplicity distribution within the pseudorapidity region -2 $<\eta<$ 0 together with Monte Carlo (MC) Glauber calculations~\cite{Miller:2007ri,STAR:2009sxc}.
The directed flow ($v_{1}$) is measured with respect to the first-order event plane, determined by the event plane detector (EPD)~\cite{Adams:2019fpo} which covers $-5.3 <\eta< -2.6$ for the FXT setup. 
For this analysis, a relatively wide centrality range, 5\%-40\%, is selected where both the event plane resolution and the hypernuclei yield are maximized. The event plane resolution in the centrality range is 40\%-75\%~\cite{STAR:2021Lan}. Detailed information on the event plane resolution can be found in the Supplemental Material~\cite{STAR:2023clhu}.

\begin{figure}[h]
\includegraphics[width=0.9\columnwidth]{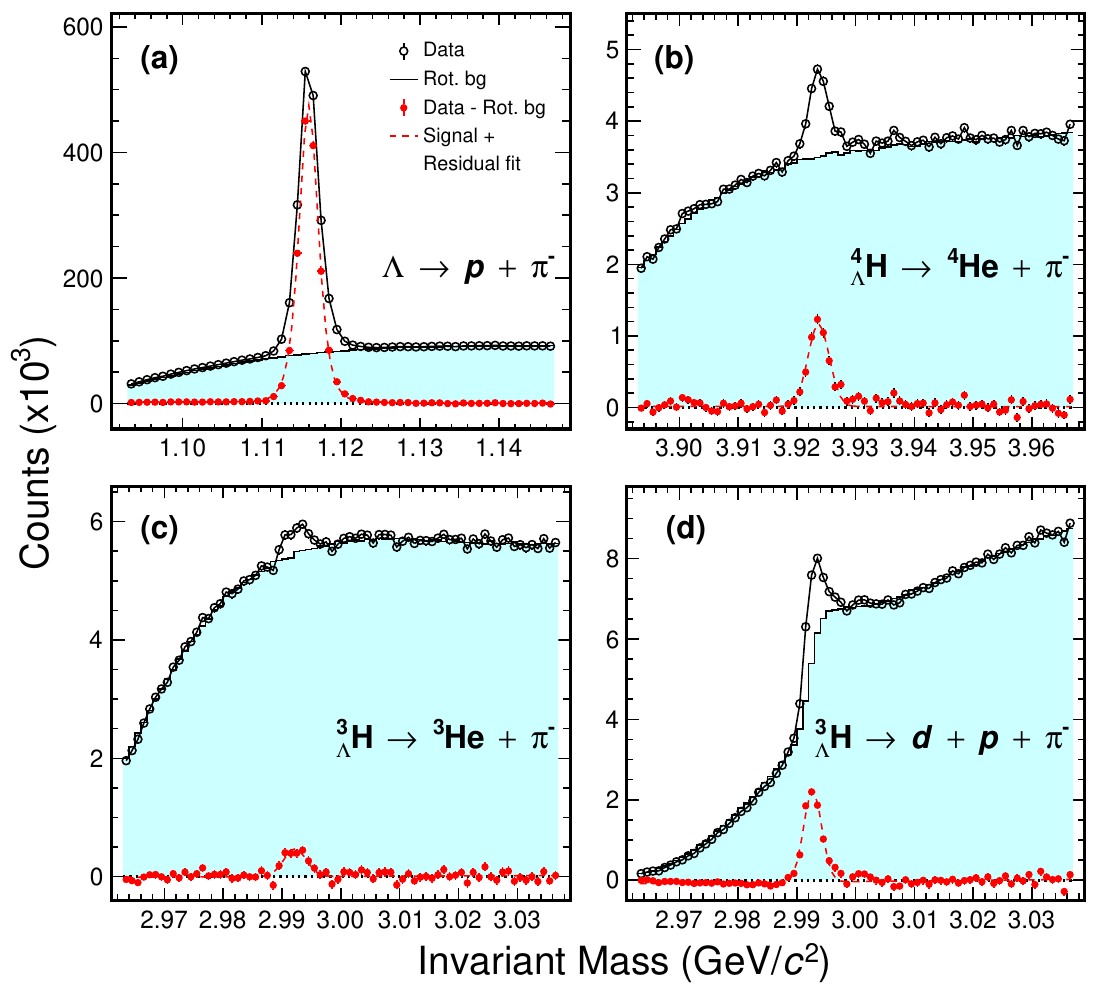}
\caption{Reconstructed $\Lambda$ hyperon and hypernuclei invariant mass distributions from \snn3 Au+Au collisions in the corresponding $p_{T}$-$y$ regions listed in Table \ref{tab1}. While top panels are for $\HL$ and $\HFTW$, bottom panels represent the hypertriton two-body decay $\HSTW$ and three-body decay $\HSTr$, respectively. Combinatorial backgrounds, shown as histograms, are constructed by rotating decay daughter particles. Background-subtracted invariant mass distributions are shown as filled circles.}
\label{fig1} 
\end{figure}

In order to ensure high track quality, we require that the number of TPC points used in the track fitting (nHitsFit) to be larger than 15 (out of a maximum of 45). $\HS$ is reconstructed via both two-body and three-body decays $\HSTW$ and $\HSTr$, while $\HF$ is reconstructed via the two-body decay channel, $\HFTW$. 
Charged particles, including $\pi^-$, $p$, $d$, $^3$He and $^4$He are selected based on the ionization energy loss ($dE$/$dx$) measured in the TPC as a function of rigidity ($p/|q|$), where $p$ and $q$ are the momentum and charge of the particle.
The secondary decay topology is reconstructed using the KFParticle package based on a Kalman filter method~\cite{Kisel:2018nvd,Zyzak:2016exl}.
The package also utilizes the covariance matrix of reconstructed tracks to construct a set of topological variables. Selection cuts on these variables are placed on hypernuclei candidates to enhance the signal significance.
Figure~\ref{fig1} shows the reconstructed invariant mass distributions for $\Lambda$, $\HS$ and $\HF$, which are reconstructed using various decay channels in the corresponding transverse momentum $p_{T}$-rapidity $y$ regions as listed in Table \ref{tab1}. Combinatorial background is estimated by rotating decay particles through a random angle between 10$^{\circ}$ and 350$^{\circ}$. For the $\Lambda$, the $\pi^-$ is rotated. For the $^{3(4)}_{\Lambda}$H two-body decay, the $\rm ^{3(4)}He$ is rotated, and for the $\HS$ three-body decay, the deuteron is rotated. The combinatorial background, shown as the shaded region, is normalized in the invariant mass region: (1.14, 1.16), (3.01, 3.04), and (3.95, 4.0) GeV/{\it{c}$^2$} for $\Lambda$, $\HS$ and $\HF$, respectively.
The background-subtracted invariant mass distribution (filled circles) in each panel is fitted with a linear function plus a student-t distribution for $\Lambda$ and a Gaussian distribution for hypernuclei to extract the signal count. In total, 8400 $\HS$ and 5200 $\HF$ reconstructed hypernuclei from the 5\%-40\% centrality bin are used for further analysis.

\begin{figure}[htb]
\includegraphics[width=0.95\columnwidth]{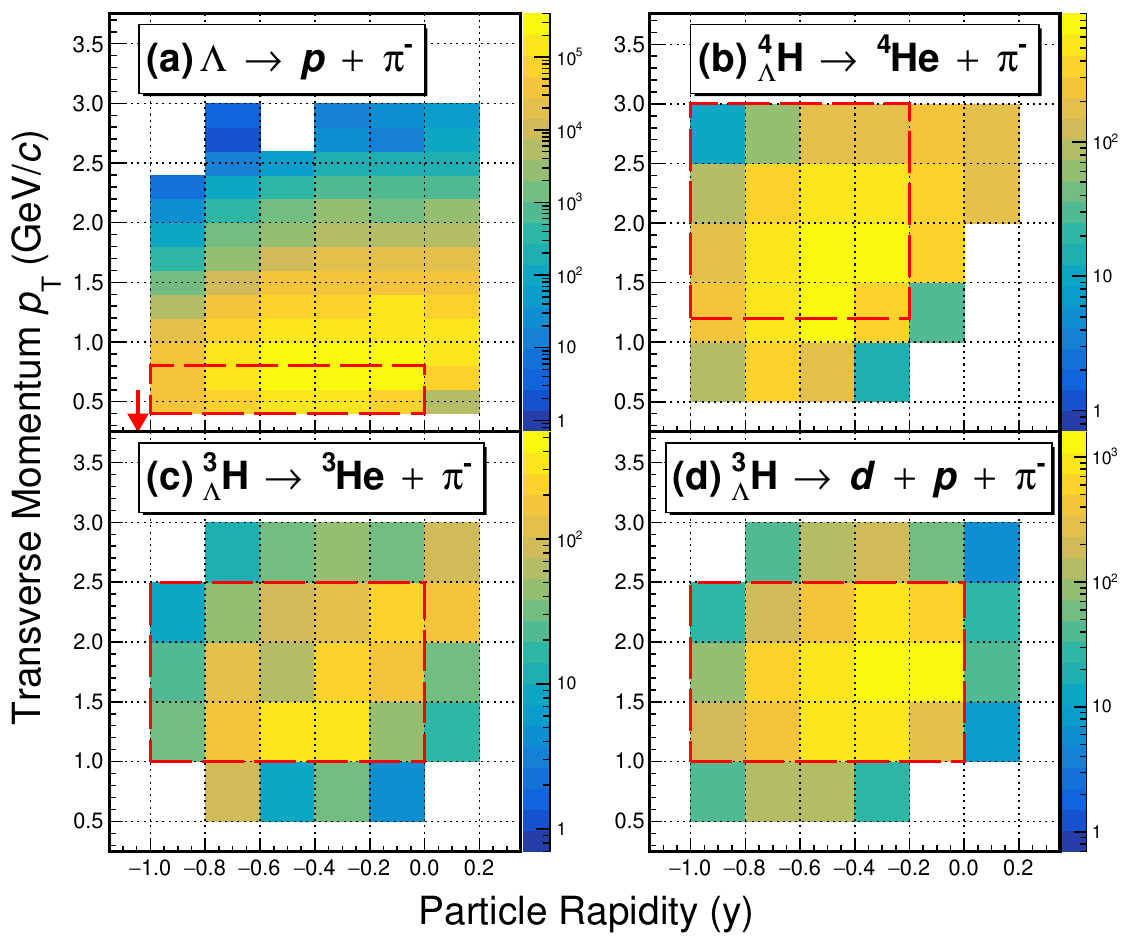}
\caption{$\Lambda$ hyperon and hypernuclei acceptance, shown in $p_{T}$ versus $y$, from the \snn3 Au+Au collisions. Dashed rectangular boxes illustrate the acceptance regions used for directed flow analysis, and the red arrow in panel (a) represents the target rapidity ($y_{\rm target}$ = $-$1.045).}
\label{fig2}
\end{figure}

\begin{table}[htb] 
    \centering
    \caption{$p_{T}$-$y$ acceptance windows of light nuclei, $\Lambda$ hyperon and hypernuclei used for directed flow analysis.} 
    \begin{tabular}{lccc} \hline \hline
    Mass number (A)    & Particle           & $p_{T}$ ({GeV/\it{c}})  & $y$ \\ \hline
    1                  & $\Lambda$, $p$     & (0.4, 0.8)   & (-1.0, 0.0)    \\ 
    2                  & $d$                & (0.8, 1.6)   & (-1.0, 0.0)    \\ 
    \multirow{2}{*}{3} & $\HS$              & (1.0, 2.5)   & (-1.0, 0.0)    \\ 
                       & $t$, $\rm ^{3}He$  & (1.2, 2.4)   & (-1.0, -0.1)   \\ 
    \multirow{2}{*}{4} & $\HF$              & (1.2, 3.0)   & \multirow{2}{*}{(-1.0, -0.2)}   \\
                       & $\rm ^{4}He$       & (1.6, 3.2)   &    \\ \hline \hline
    \end{tabular}
    \label{tab1}
\end{table}

Figure~\ref{fig2} shows the $p_{T}$ versus $y$ acceptance of the reconstructed $\Lambda$, $\HS$, and $\HF$ candidates in the center-of-mass frame. Following the established convention~\cite{E895:2000maf}, the negative sign is assigned to $v_1$ in the rapidity region of $y <$ 0. The $p_{T}$-$y$ acceptance windows used for our analysis are tabulated in Table~\ref{tab1} and also indicated in Fig.~\ref{fig2}.

For $p_{T}$-integrated $v_1$ measurements, the $p_{T}$-dependent reconstruction efficiency needs to be accounted for, which is estimated by the embedding method in STAR analyses~\cite{STAR:2019bjj,STAR:2021orx}. Monte Carlo generated hyperons and hypernuclei are passed through the GEANT3 simulation of the STAR detector. The simulated TPC response is then embedded into data, and the whole event is processed and analyzed using the same procedure as in the data analysis. The two-dimensional reconstruction efficiency, including the detector acceptance, in $p_{T}$-$y$ are obtained for each decay channel, and applied to candidates in the data accordingly~\cite{STAR:2021yiu}. Kinematically, the three-body decay of $\HS$ is very similar to the background of correlated $d + \Lambda$ due to the very small $\Lambda$ separation energy of $\HS$. Such correlated $d + \Lambda$ pairs that pass the $\HS$ three-body decay topological cuts are subtracted statistically (For details, see Fig. 3 in the Supplemental Material~\cite{STAR:2023clhu}, which includes~\cite{Haidenbauer:2020uew}). The $\HS$ signal fraction within the invariant mass window (2.988, 2.998) {GeV/\it{c}$^2$} and rapidity range (-1.0, 0.0) is estimated to be $0.69 \pm 0.03$. 

The directed flow of $\Lambda$, $\HS$, and $\HF$ are extracted with the event plane method~\cite{Masui:2012zh}. In each rapidity bin, the azimuthal angle with respect to the reconstructed event plane (${\Phi}={\Phi}'-\Psi_1$) is further divided into four equal bins with a width of $\pi/4$, where $\Phi'$ and $\Psi_1$ are the azimuth angle of a particle candidate and the first order event plane, respectively. After applying the reconstruction efficiency correction, the azimuthal angle distributions are fitted with a function $f({\Phi})=c_0[1 + 2v_1^{\rm obs}{\cdot}\cos(\Phi) + 2v_2^{\rm obs}{\cdot}\cos(2\Phi)]$, where $c_0$, $v_1^{\rm obs}$ and $v_2^{\rm obs}$ are fitting parameters, and correspond to the normalization constant, the observed directed and the elliptic flow, respectively. 
To obtain the final $v_1$ in a wide centrality range of 5\%-40\% centrality in this analysis, the observed directed flow $v_1^{\rm obs}$ needs to be corrected for the average event plane resolution $\langle 1/R\rangle$~\cite{Masui:2012zh}, i.e., $v_1=v_1^{\rm obs}\cdot\langle 1/R\rangle$, and $\langle 1/R \rangle = \sum_i{(N_i/R_i)}/\sum_i{N_i}$, where $N_{i}$ and $R_i$ stand for the number of particle candidates and the first order event plane resolution in the $i$th centrality bin, respectively. 

\begin{figure}[htb]
\includegraphics[width=0.9\columnwidth]{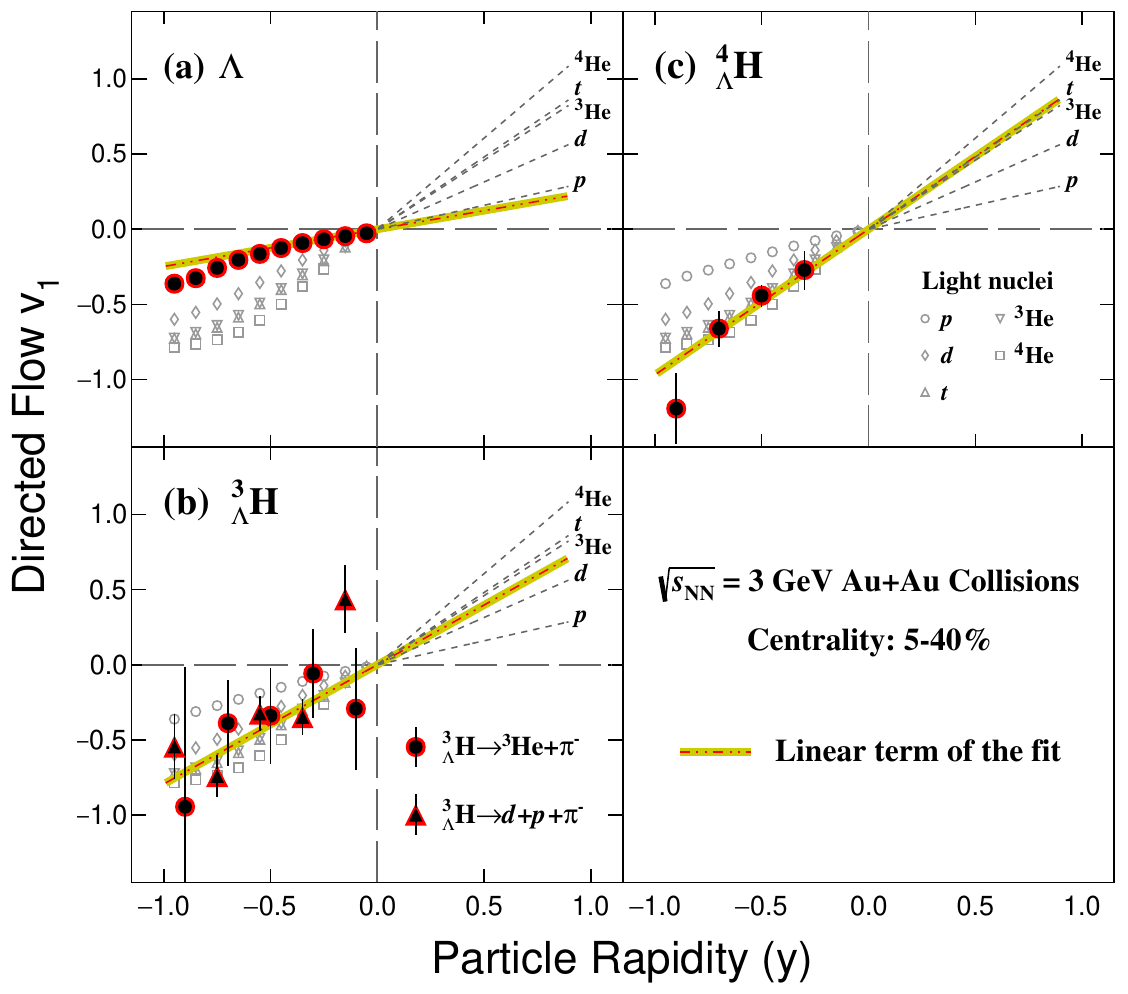}
\caption{
$\Lambda$ hyperon and hypernuclei directed flow $v_1$, shown as a function of rapidity, from the \snn3 5\%-40\% midcentral Au+Au collisions. In the case of $\HS$ $v_1$, both two-body (dots) and three-body (triangles) decays are used. The linear terms of the fitting for $\Lambda$, $\HS$ and $\HF$ are shown as the yellow-red lines. The rapidity dependence of $v_1$ for $p$, $d$, $t$, $^3$He, and $^4$He are also shown as open markers (circles, diamonds, up triangles, down triangles and squares), and the linear terms of the fitting results are shown as dashed lines in the positive rapidity region~\cite{STAR:2021ozh}. }
\label{fig3}
\end{figure}

The resulting $\Lambda$ hyperon and hypernuclei $v_1(y)$, from 5\%-40\% midcentral Au+Au collisions at \snn3 , are shown in Fig.~\ref{fig3}. For comparison, the $v_1(y)$ of $p$, $d$, $t$, $^3\rm{He}$ and $^4\rm{He}$ from the same data~\cite{STAR:2021ozh} are shown as open symbols. 
$v_1(y)$ of $\Lambda$, $p$, $d$, $t$, $^3$He and $^4$He are fitted with a third-order polynomial function $v_1(y)$ = $ay$ + $by^3$ in the rapidity ranges listed in Table \ref{tab1}, where $a$, which stands for the midrapidity slope $dv_1/dy|_{y=0}$, and $b$ are fitting parameters. 
Because of limited statistics, the hypernuclei $v_1(y)$ distributions are fitted with a linear function $v_1(y)$ = $ay$, in the rapidity range $-1.0<y<0.0$. The linear terms for light nuclei are plotted as dashed lines in the positive rapidity region, while for $\Lambda$, $\HS$, and $\HF$, they are shown by the yellow-red lines in the corresponding panels.
The $\Lambda$ result is close to that of the proton, and hypernuclei $v_1(y)$ distributions are also similar to those light nuclei with the same mass numbers. This is the first observation of significant hypernuclei directed flow in high-energy nuclear collisions.

Systematic uncertainties are estimated by varying track selection criteria for particle identification, as well as cuts on the topological variables used in the KFParticle package~\cite{Kisel:2018nvd}. Major contributors to the systematic uncertainty are listed in Table~\ref{tab2}. As one can see, the dominant sources of systematic uncertainty are from hypernuclei candidate selection, estimated by varying topological cuts and nHitsFit. 
Event plane resolution determination also contributes 1.4\%~\cite{STAR:2021yiu}. Assuming these sources are uncorrelated, the total systematic uncertainty is obtained by adding them together quadratically. In the case of the $\HS$ three-body decay, the fraction of the correlated {\it{d}$\Lambda$} contamination has been analyzed in each rapidity bin. Its systematic uncertainty contribution to the final $v_1$ slope is negligible. 


\begin{table}[htb]
\centering
\caption{Sources of systematic uncertainties for midrapidity slope $dv_1/dy|_{y=0}$ of $\HS$ and $\HF$.}
\begin{tabular}{lccc}
\hline \hline
                 & \multicolumn{2}{c}{$\HS$}                 & $\HF$            \\ \cline{2-4}
Source           & \multicolumn{1}{c}{two-body} & three-body & two-body         \\ \hline
Topological cuts & \multicolumn{1}{c}{1.3\%}    & 9.4\%      & 8.0\%            \\ 
nHitsFit         & \multicolumn{2}{c}{9.0\%}                 & $<$1.0\%         \\ 
EP resolution    & \multicolumn{2}{c}{1.4\%}                 & 1.4\%            \\ 
Total            & \multicolumn{2}{c}{13.1\%}                & 8.3\%            \\ \hline \hline
\end{tabular}
\label{tab2}
\end{table}

\begin{figure}[htb]
\includegraphics[width=0.9\columnwidth]{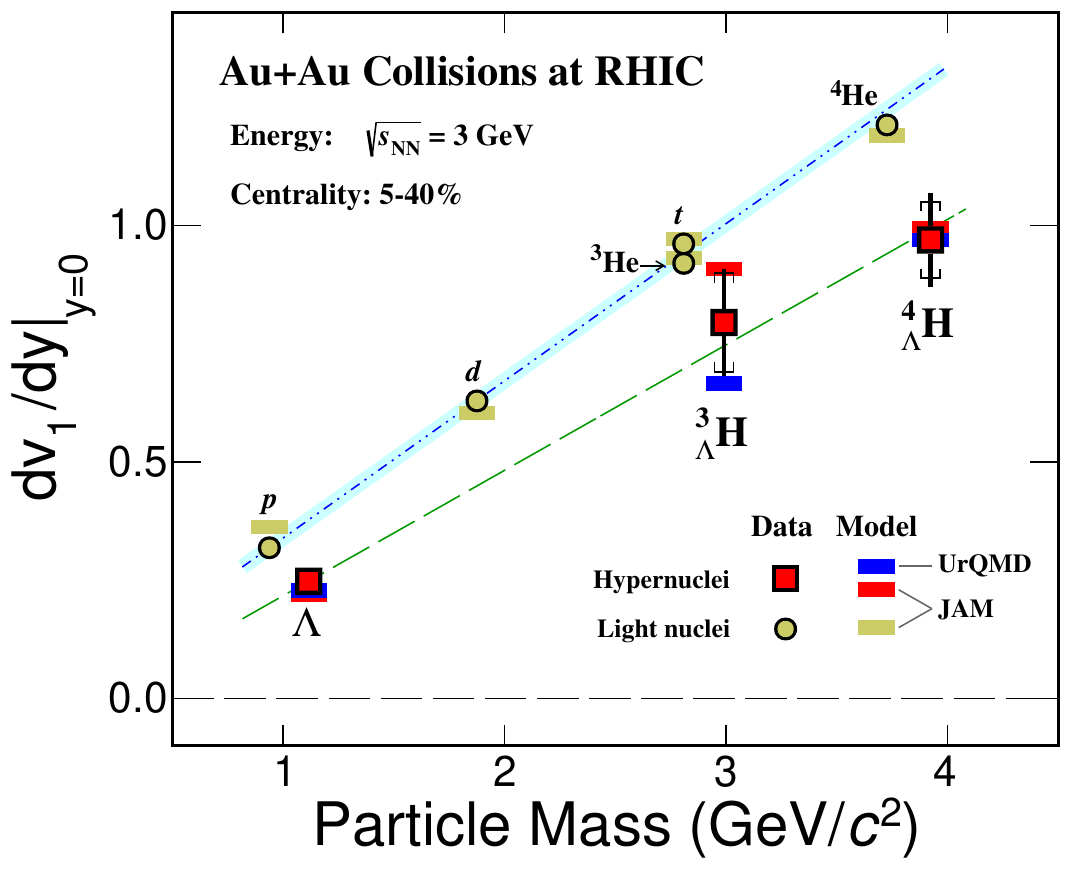}
\caption{Mass dependence of the midrapidity $v_1$ slope, $dv_1/dy$, for $\Lambda$, $\HS$ and $\HF$ from the \snn3 5\%-40\% midcentral Au+Au collisions. The statistical and systematic uncertainties are presented by vertical lines and square brackets, respectively. The slopes of $p$, $d$, $t$, $^3$He, and $^4$He from the same collisions are shown as black circles. The blue and dashed green lines are the results of a linear fit to the measured light nuclei and hypernuclei $v_1$ slopes, respectively. For comparison, calculations of transport models plus coalescence afterburner are shown as gold and red bars from the JAM model, and blue bars from the UrQMD model.}
\label{fig4}
\end{figure}

The results of the midrapidity slope $dv_1/dy$ for $\Lambda$, $\HS$ (both two- and three-body decays) and $\HF$ are shown in Fig.~\ref{fig4}, as filled squares, as a function of particle mass. For comparison, $v_1$ slopes of $p$, $d$, $t$, $^3$He and $^4$He from the same 5\%-40\% \snn3 Au+Au collisions are shown as open circles. The $\Lambda$ hyperon and hypernuclei slopes $dv_1/dy$ are all systematically lower than the nuclei of same mass numbers. 
Linear fits ($f = a + b$ mass) are performed on the mass dependence of $dv_1/dy$ for both light nuclei and hypernuclei. For light nuclei, only statistical uncertainties are used in the fit, while statistical and systematic uncertainties are used for hypernuclei. The slope parameters $b$ are $0.3323 \pm 0.0003$ for light nuclei and $0.27 \pm 0.04$ for hypernuclei. As one can see, their slopes are similar within uncertainties.

Using transport models JAM~\cite{Nara:2016phs,Nara:1999dz} and UrQMD~\cite{Steinheimer:2014pfa}, $v_1(y)$ of $\Lambda$ and hypernuclei are simulated for the \snn3 Au+Au collisions within the same centrality and kinematic acceptance used in data analysis. For comparison, similar calculations are performed for light nuclei. The simulation is done in two steps: (i) using the JAM model (with momentum-dependent potential) and UrQMD model (without momentum-dependent potential) in the mean field mode with the incompressibility $\kappa=380$ MeV to produce neutrons, protons, and $\Lambda$s at kinetic freeze-out; (ii) forming hypernuclei through the coalescence of $\Lambda$ and nucleons, similar to the light nuclei production with the coalescence procedure discussed in~\cite{STAR:2021ozh}. The probability for hypernuclei production is dictated by coalescence parameters of relative momenta $\Delta p < 0.12$ (0.3)\,GeV/it{c} and relative distance $\Delta r < 4$\,fm in the rest frame of $np\Lambda$ ($nnp\Lambda$) for $\HS$($\HF$).
These parameters are chosen such that the hypernuclei yields at midrapidity can be described~\cite{STAR:2021orx}. The rapidity dependences of $v_1$ from the model calculations are then fitted with a third-order polynomial function within the rapidity interval $-1.0\leq y \leq 0.0$. The resulting midrapidity slopes are shown in Fig.~\ref{fig4} as red and blue bars for JAM and UrQMD models, respectively. In the figure, results for light nuclei from JAM are also presented as gold bars.

Both transport models (JAM and UrQMD) plus coalescence afterburner calculations for hypernuclei are in agreement with data within uncertainties. Interactions among baryons and strange baryons are important ingredients in the transport models, especially in the high baryon density region~\cite{Botvina:2016wko,Botvina:2014lga}. The properties of the medium are determined by such interactions. In addition, the yields of hypernuclei, if created via the coalescence process, are also strongly affected by the hyperon and nucleon interactions. 
In our treatment, the coalescence parameters used ($\Delta r$, $\Delta p$) reflect the production probability determined by $N$-$N$ and $Y$-$N$ interactions~\cite{Shao:2020lbq,Wang:1999bf,Aichelin:2019tnk}. 
The mass dependence of the $v_1(y)$ slope implies that coalescence might be the dominant mechanism for hypernuclei production in such heavy-ion collisions. The mass dependence of the hypernuclei $v_1$ slope also seems to be similar to that of light nuclei, as shown in Fig.~\ref{fig4}, although it may not necessarily be so due to the differences in $N$-$N$ and $Y$-$N$ interactions. Clearly, precision data on hypernuclei collectivity will yield invaluable insights on $Y$-$N$ interactions at high baryon density.

This is the first report of the collectivity of hypernuclei in heavy-ion collisions. Hydrodynamically, collective motion is driven by pressure gradients created in such collisions. This letter opens up a new direction for studying $Y$-$N$ interaction under finite pressure~\cite{Neidig:2021bal}. This is important for making the connection between nuclear collisions and the equation of state which governs the inner structure of compact stars.

To summarize, we report the first observation of hypernuclei $\HS$ and $\HF$ $v_1$ from \snn3 midcentral 5\%-40\% Au+Au collisions at RHIC. The rapidity dependences of their $v_1$ are compared with those of $\Lambda$, $p$, $d$, $t$, $^3$He and $^4$He in the same collisions. It is found that, within uncertainties, the mass dependent $v_1$ slope of hypernuclei, $\HS$ and $\HF$ is similar to that of light nuclei, implying that they follow the baryon mass scaling. Calculations from transport models (JAM and UrQMD) plus coalescence afterburner can qualitatively reproduce the rapidity dependence of $v_1$ and the mass dependence of the $v_1$ slope. These observations suggest that coalescence of nucleons and hyperon $\Lambda$ could be the dominant mechanism for the hypernuclei $\HS$ and $\HF$ production in the 3 GeV collisions. 
Model calculations suggest that baryon density at freeze-out may depend on collision energy~\cite{Reichert:2020yhx,Cleymans:2005xv, Randrup:2006nr}. High statistics data at different energies, especially at the high baryon density region, will help in extracting the information on $Y$-$N$ interaction and possibly its density dependence in the future.

We thank Dr. Y. Nara and Dr. J. Steinheimer for insightful discussions. 
We thank the RHIC Operations Group and RCF at BNL, the NERSC Center at LBNL, and the Open Science Grid consortium for providing resources and support.  This work was supported in part by the Office of Nuclear Physics within the U.S. DOE Office of Science, the U.S. National Science Foundation, National Natural Science Foundation of China, Chinese Academy of Science, the Ministry of Science and Technology of China and the Chinese Ministry of Education, the Higher Education Sprout Project by Ministry of Education at NCKU, the National Research Foundation of Korea, Czech Science Foundation and Ministry of Education, Youth and Sports of the Czech Republic, Hungarian National Research, Development and Innovation Office, New National Excellency Programme of the Hungarian Ministry of Human Capacities, Department of Atomic Energy and Department of Science and Technology of the Government of India, the National Science Centre and WUT ID-UB of Poland, the Ministry of Science, Education and Sports of the Republic of Croatia, German Bundesministerium f\"ur Bildung, Wissenschaft, Forschung and Technologie (BMBF), Helmholtz Association, Ministry of Education, Culture, Sports, Science, and Technology (MEXT) and Japan Society for the Promotion of Science (JSPS).

\bibliography{hyper_flow} 

\end{document}


\title{Supplemental Material: 
First Observation of Directed Flow of hypernuclei $^3_{\Lambda}$H and $^4_{\Lambda}$H in \snn3 Au+Au Collisions at RHIC} 
\author{The STAR Collaboration}

\maketitle 

\section{Event plane resolution}

\begin{figure}[htb]
\centering
\includegraphics[width=0.5\columnwidth]{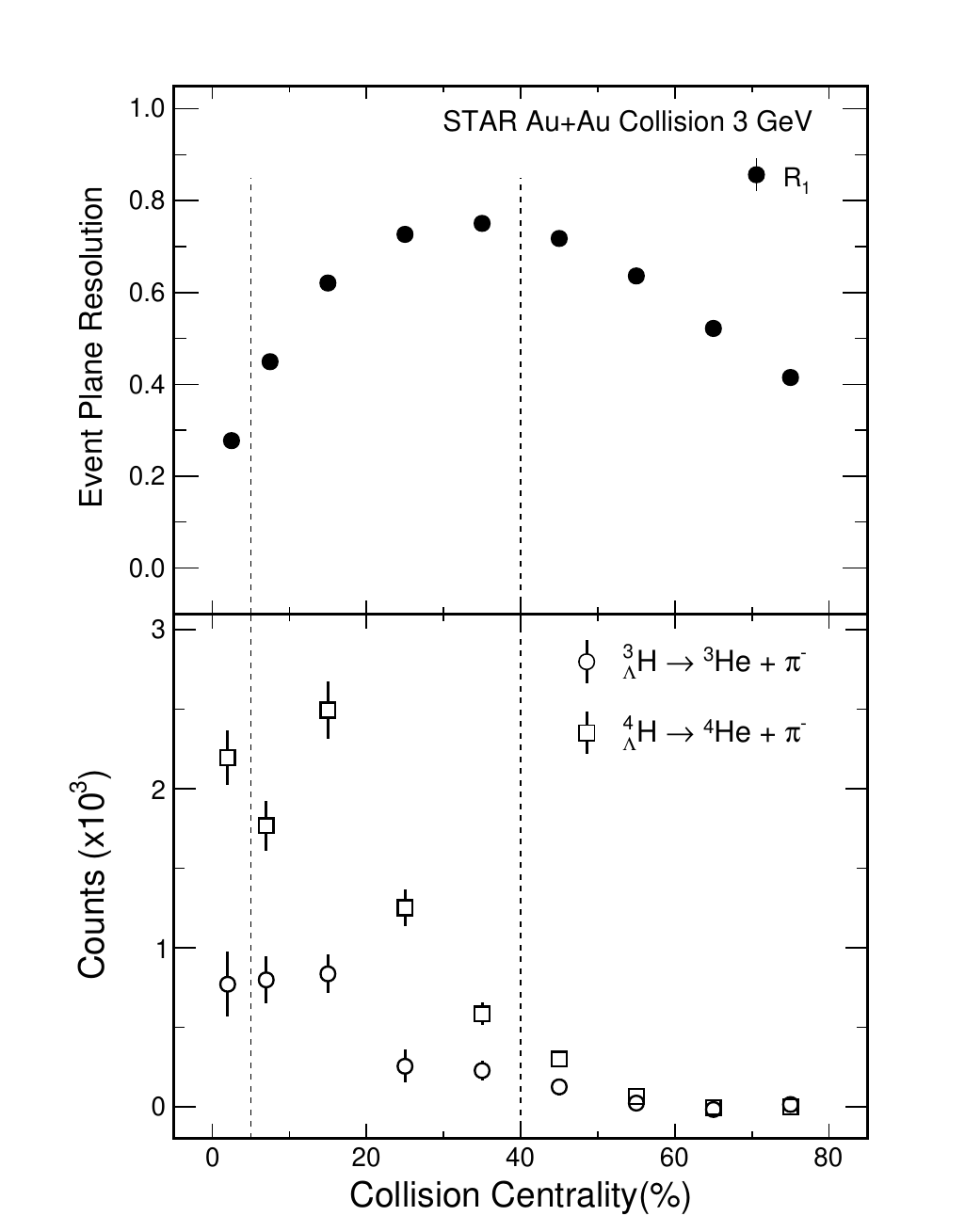}
\caption{Top: First-order event plane resolution as a function of collision centrality. Bottom: Distribution of raw yields as a function of collision centrality for $\rm ^{3}_{\Lambda}H$ 2-body decay and $\rm ^{4}_{\Lambda}H$.}
\label{fig_EPR_cent}
\end{figure}

In this measurement, the directed flow is calculated via the event-plane method \cite{Masui:2016EP}. For each event, the event plane is reconstructed using the Event Plane Detector (EPD)~\cite{Adams:2019fpo}. For the estimation of the resolution for the first-order event plane, please refer to~\cite{STAR:2021Lan} for details. Figure~\ref{fig_EPR_cent} shows the first-order event plane resolution (Top) and the distributions of raw yields for $\rm ^{3}_{\Lambda}H$ 2-body decay and $\rm ^{4}_{\Lambda}H$ as a function of collision centrality (Bottom). Events from centrality 5-40\% (shown by the black dashed lines) are selected for analysis to optimize the signal. 

\section{Topological cuts used for $\Lambda$, $^{3}_{\Lambda}$H and $^{4}_{\Lambda}$H}

Table~\ref{tab_topo} shows topological variables cuts employed for $\Lambda$, $\rm ^{3}_{\Lambda}H$, and $\rm ^{4}_{\Lambda}H$ candidate selection. Their definitions are: (1) nHitsFit is the number of TPC points used in the track fitting; (2) $l$ is the decay length from primary vertex to decay vertex; (3) $ldl$ is the distance from the decay point of the candidate to the primary vertex normalized on the error; (4) $\chi^{2}_{topo}$ is the mother particle's $\chi^{2}$ deviation from the primary vertex; (5) $\chi^{2}_{ndf}$ is the $\chi^{2}$ deviation between daughter particles; (6) $\chi^{2}_{prim}$ defines the daughter particle's $\chi^{2}$ deviation from the primary vertex.

\begin{table}[htb]
\begin{tabular}{c|c|c|c|c}
\hline
Particle  & $\Lambda$   & $\rm ^{3}_{\Lambda}H$ (2-body)  & $\rm ^{3}_{\Lambda}H$ (3-body)   & $\rm ^{4}_{\Lambda}H$ \\ \hline
\multirow{8}{*}{Topological cuts} & nHitsFit $\geq$ 15           & nHitsFit $\geq$ 15   & nHitsFit $\geq$ 15      & nHitsFit $\geq$ 15           \\
 & $l$ \textgreater{} 1 cm  & $l$ \textgreater{} 2 cm  & $l$ \textgreater{} 8 cm  & $l$ \textgreater{} 1.5 cm                \\
 & $ldl$ \textgreater{} 5   & $ldl$ \textgreater{} 3   & $ldl$ \textgreater{} 5   & $ldl$ \textgreater{} 3                   \\
 & $\chi^{2}_{topo}$ $<$ 5  & $\chi^{2}_{topo}$ $<$ 5  & $\chi^{2}_{topo}$ $<$ 3   & $\chi^{2}_{topo}$ $<$ 3                 \\
 & $\chi^{2}_{ndf}$ $<$ 5       & $\chi^{2}_{ndf}$ $<$ 5       & $\chi^{2}_{ndf}$ $<$ 3.5     & $\chi^{2}_{ndf}$ $<$ 2       \\
 & $\chi^{2}_{prim,p}$ $>$ 10   & $\chi^{2}_{prim,He}$ $>$ 5   & $\chi^{2}_{prim,d}$ $>$ 0    & $\chi^{2}_{prim,He}$ $>$ 0   \\
 & $\chi^{2}_{prim,\pi}$ $>$ 10 & $\chi^{2}_{prim,\pi}$ $>$ 20 & $\chi^{2}_{prim,p}$ $>$ 5    & $\chi^{2}_{prim,\pi}$ $>$ 10 \\
 &                              &                              & $\chi^{2}_{prim,\pi}$ $>$ 10 &                              \\ \hline
\end{tabular}
\caption{Topological cuts used for $\Lambda$, $^{3}_{\Lambda}{\rm H}$ and $ ^{4}_{\Lambda}{\rm H}$.}
\label{tab_topo}
\end{table}

\section{FITTING AZIMUTHAL DISTRIBUTIONS FOR HYPERNUCLEI}

\begin{figure}[htb]
	\centering
    \includegraphics[width=0.96\columnwidth]{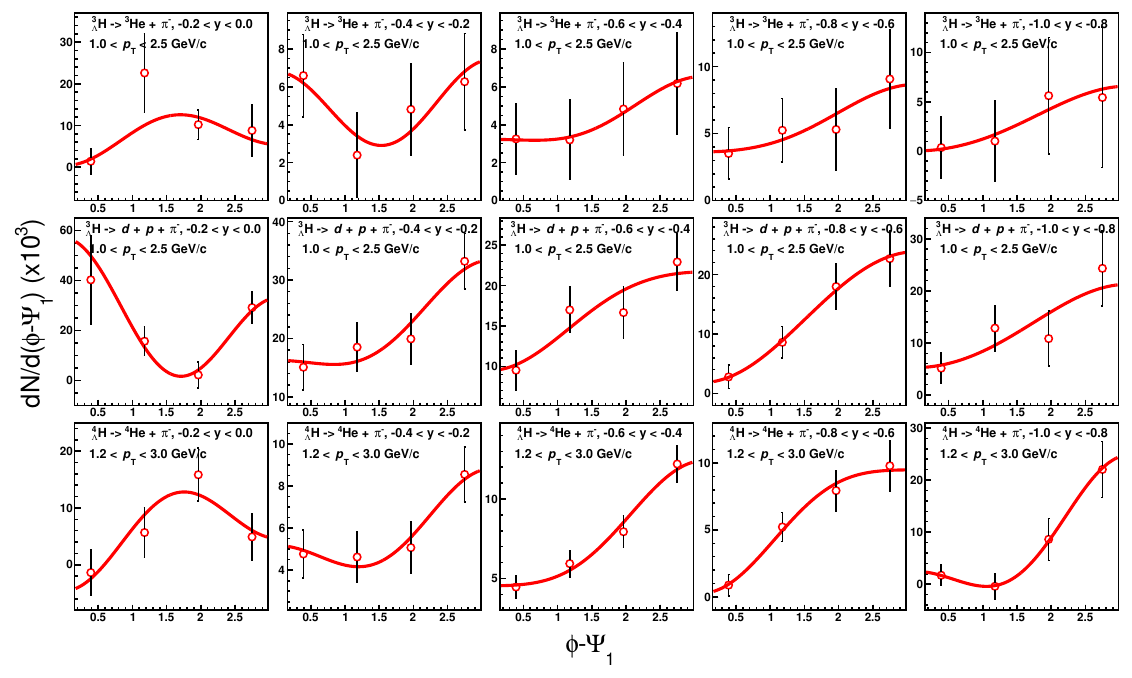}
    \caption{In 3 GeV Au+Au collisions, the angular distributions of $\rm ^{3}_{\Lambda}H$ 2-body and 3-body decays and $\rm ^{4}_{\Lambda}H$ in different rapidity regions.}
	\label{fig_dNdphi}
\end{figure}

Figure \ref{fig_dNdphi} shows the distributions of $\rm dN/d(\phi-\Psi_1$) as a function of $\phi - \Psi_1$ for $\rm ^{3}_{\Lambda}H$ 2-body and 3-body decay and $\rm ^{4}_{\Lambda}H$ in different rapidity regions, for centrality bin 5-40\%. Distributions of $\rm dN/d(\phi-\Psi_1$) is fitted by $\rm dN/d(\phi-\Psi_1) = p_0(1+2v_1^{obs}cos(\phi-\Psi_1)+2v_2^{obs}cos(2(\phi-\Psi_1)))$, where $\rm p_0$, v$_1^{obs}$ and v$_2^{obs}$ are normalization parameter, observed directed and elliptic flows, respectively. In each panel, red lines show the fitting results, which can well describe data points. 

\section{Purity for the $\HS$ $\rightarrow p+d+\pi^{-}$ 3-body decay}
The kinematically correlated $\Lambda$ and deuteron would form a peak around $M(\Lambda)+M(d)=$ 2.9913 GeV/$c$ when $C(k^{*})>1$ at $k^{*}\rightarrow 0$, where $C(k^{*})$ is the kenimatic correlation function of $\Lambda$ and deuteron and $k^{*}$ is relative momentum between $\Lambda$ and deuteron. These correlated $\Lambda+d$ pairs would coincide with the reconstructed $\HS$ signals via $\HS$ $\rightarrow p+d+\pi^{-}$ channel even after combinatorial background subtraction, since $\HS$ is weakly bounded with a small $\Lambda$ separate energy $B_{\Lambda}\sim$ 0.13 - 0.41 MeV/c$^2$ \cite{STAR:2019wjm} and the finite experimental momentum resolution could not separate the $\HS$ invariant mass peak from the correlated $\Lambda+d$ peak.

\begin{figure}[htb]
	\centering
	\includegraphics[width=0.5\columnwidth]{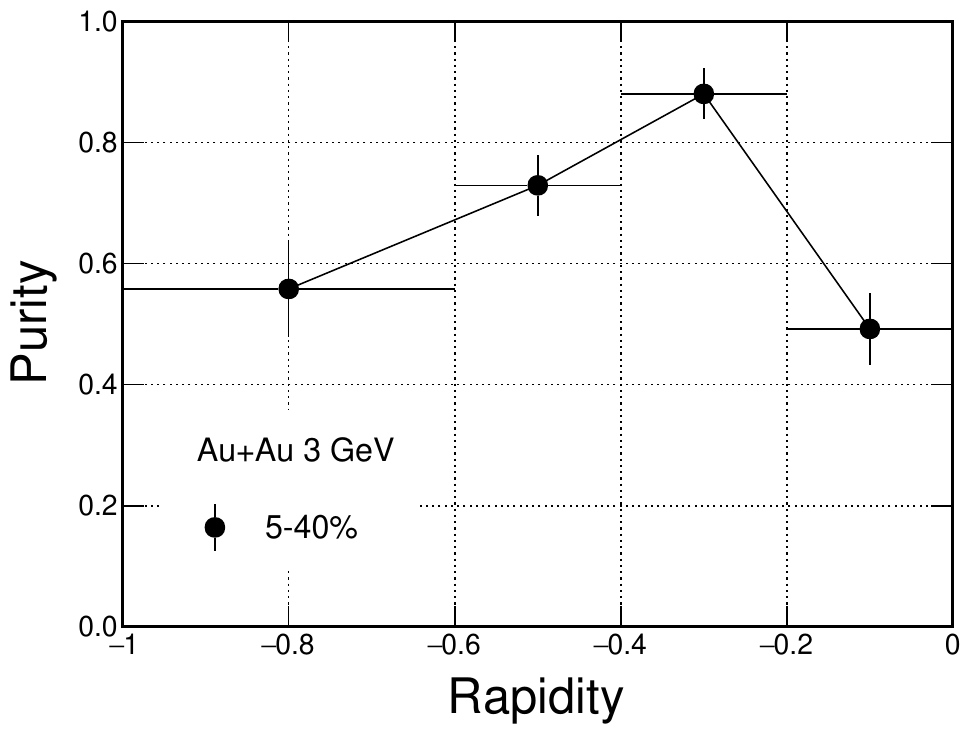}
	\caption{The estimated purity of $\HS$ $p_{\rm ^3_{\Lambda}H}$ in the reconstructed $\HS$ candidate sample via $\HS$ $\rightarrow p+d+\pi^{-}$ as a function of rapidity.}
	\label{fig:purity}
\end{figure}

Although the invariant masses of correlated $\Lambda+d$ background and the real signal are close, some of their topological kinematics are different.  For correlated $\Lambda+d$ background, the decay daughters of $\Lambda\rightarrow p\pi^{-}$ and deutreon come from different vertices, while the daughters of $\HS$ are all produced at the same vertex. $\chi^{2}_{NDF}$ is a topological variable calculated by KFParticle package \cite{Zyzak:2016exl} that characterizes whether particle trajectories intersect at the same vertex within uncertainties. A cut on $\chi^{2}_{NDF}$ can greatly suppress the correlated background. However, these correlated background cannot be fully rejected by such a cut, since $\Lambda$ daughters and the deuteron could be very close at $k^{*}\rightarrow 0$ and they cannot be distinguished experimentally due to the finite spatial resolution. Therefore, the template fitting method is used to extract the fraction of $\HS$ signal in the reconstructed sample. $\chi^{2}_{NDF}$ distributions of the candidate $\HS$ in the data are extracted and then fitted with the template $\chi^{2}_{NDF}$ distributions of correlated background and pure signal. The $\HS$ candidates from data are selected within $M(\HS)\pm2\sigma$ with all topological cuts used in the analysis applied expect for the $\chi^{2}_{NDF}$ cut. The combinatorial background is estimated by rotating decay daughter particles. The templates for $\HS$ signals are obtained from embedding simulated Monte Carlo (MC) $\HS$ signal into real data and applying the same reconstruction procedure as in data analysis. Similarly, correlated background templates are built by embedding simulated $\Lambda$ particles into real data and pairing MC $\Lambda$ with deuteron tracks from real data. MC $\Lambda$ are weighted according to the measured $\Lambda$ spectra. The reconstructed $\Lambda+d$ from embedding are weighted with kinematic correlation functions \cite{Haidenbauer:2020uew} and applied with the same topological and invariant mass selection cuts as data samples. 
The $\chi^2_{NDF}$ distribution of reconstructed $\HS$  candidates are fitted with $f_{\rm data}=p_0\cdot(p_1\cdot f_{\hyt}+f_{\rm \Lambda d})$, where $f_{\rm data}$, $f_{\hyt}$ and $f_{\rm \Lambda d}$ refer to the normalized $\chi^2_{NDF}$ distributions of H3L candidates from real data, H3L and $\Lambda+d$ templates from embedding, respectively. The parameter $p_0$ and $p_1$ are fitting parameters. The $\HS$ purity, $p_{\rm ^3_{\Lambda}H}$, is defined as the fraction of $\HS$ signal in the reconstructed $\HS$ candidates with all topological cuts applied in the data analysis. Figure \ref{fig:purity} shows the estimated $p_{\rm ^3_{\Lambda}H}$ as a function of rapidity. The $\HS$ yield is calculated as $p_{\rm ^3_{\Lambda}H}\cdot N^{raw}$, where $N^{raw}$ is the raw $\hyt$ candidate counts. 

\section{Model analysis and comparison}

\subsection{Determine the coalescence parameters from measured hypernuclei $dN/p_{\rm T}dydp_{\rm T}$}

\begin{figure}[htb]
\includegraphics[width=0.49\columnwidth]{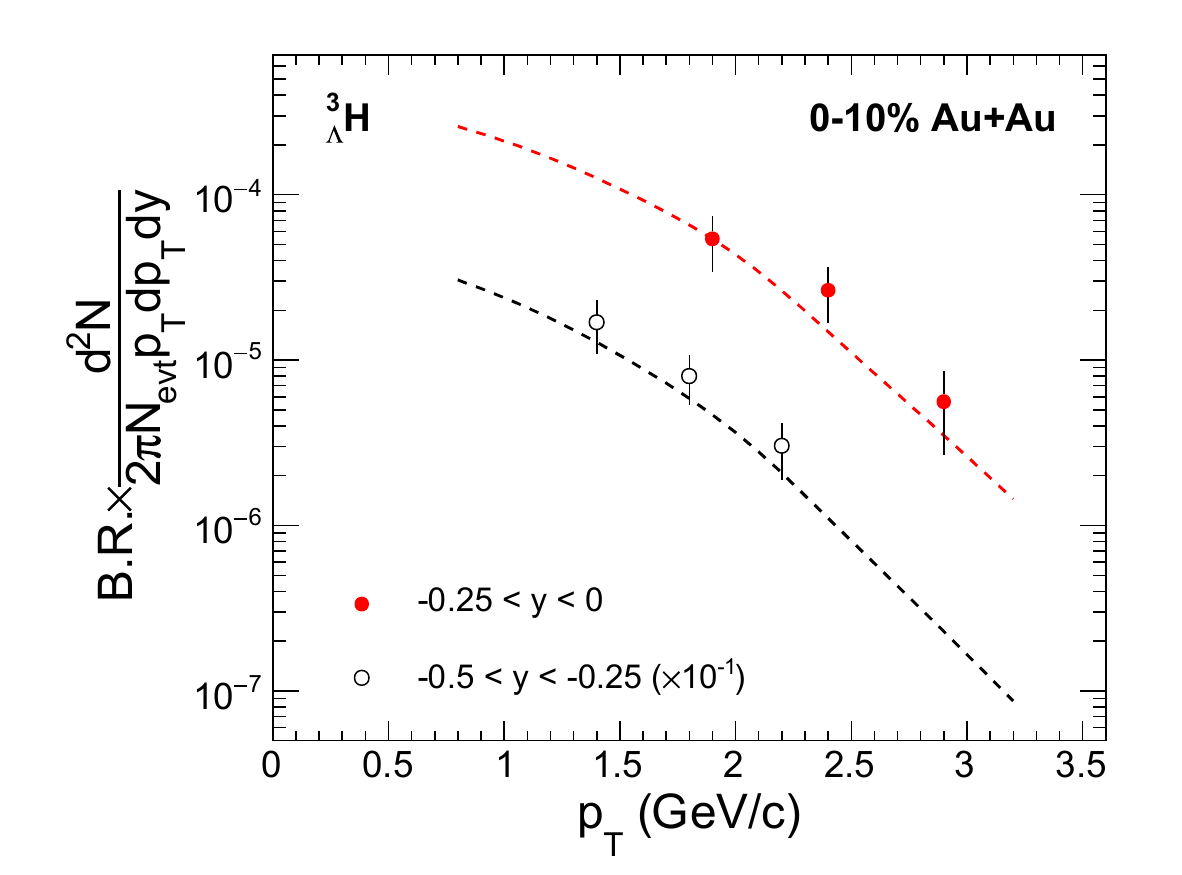} 
\includegraphics[width=0.49\columnwidth]{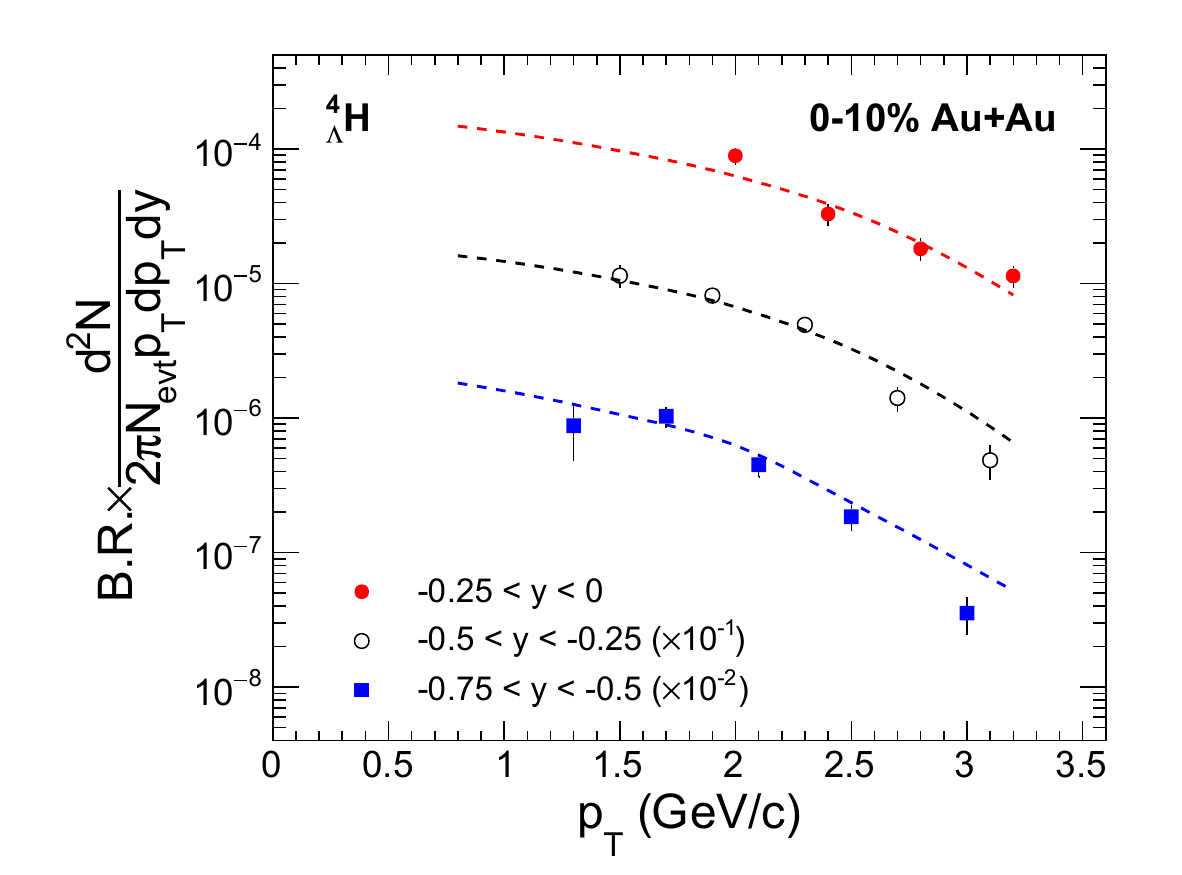} 
\caption{$\HS$ and $\HF$ $p_{\rm{T}}$ spectra in three different rapidity intervals from 0-10\% central Au+Au collisions at $\sqrt{s_{NN}}=3$ GeV. Only the statistical uncertainties are plotted. The dashed lines are results from JAM plus coalescence calculations. For model results, spectra are scaled with the branching ratios which are assumed to be 25\% and 50\% for $\HSTW$ and $\HFTW$, respectively~\cite{STAR:2021orx}.}
\label{fig:hyper_spectra}
\end{figure}

Transport models JAM~\cite{Nara:1999dz,Nara:2016phs} and UrQMD~\cite{Steinheimer:2014pfa} are utilized to generate nucleons and $\Lambda$ hyperons at 3 GeV.
As these models can not reproduce light nuclei and hypernuclei yields, a simple coalescence approach is employed to form light nuclei and hypernuclei using spatial and momentum distributions of nucleons and $\Lambda$ hyperons at a time of 50 fm/$c$ in the medium evolution. For a hypernucleus, it forms in two steps. Firstly, a light nucleus core is formed based on the relative momentum $\Delta p$ and relative distance $\Delta r$ of constituent nucleons in their rest frame. Then the light nucleus core combines with a $\Lambda$ to form a hypernucleus following a similar process. The coalescence parameters, $\Delta p$ and $\Delta r$, are determined by matching the $dN/p_{\rm T}dydp_{\rm T}$ spectrum from the calculations to the corresponding experimental data for a given light nucleus or hypernucleus~\cite{STAR:2021orx}. The $\Delta p$ is 0.3 GeV/$c$ for both deuteron and triton, and $\Delta r$ is 4.5 fm and 4 fm for them, respectively. The $\Delta r$ is 4 fm for both $\HS$ (deuteron $+$ $\Lambda$) and $\HF$ (triton $+$ $\Lambda$), and $\Delta p$ is 0.12 GeV/$c$ and 0.3 GeV/$c$, respectively. In the model, we assume that the branching ratios are 25\% and 50\% for $\HSTW$ and $\HFTW$, respectively. These results from JAM plus coalescence calculations can qualitatively reproduce the $\HS$ and $\HF$ spectra~\cite{STAR:2021orx}, as show in Fig.~\ref{fig:hyper_spectra}. 

\subsection{$v_1(\rm{y})$ distributions from JAM and UrQMD model calculations}

\begin{figure}[htb]
\includegraphics[width=0.75\columnwidth]{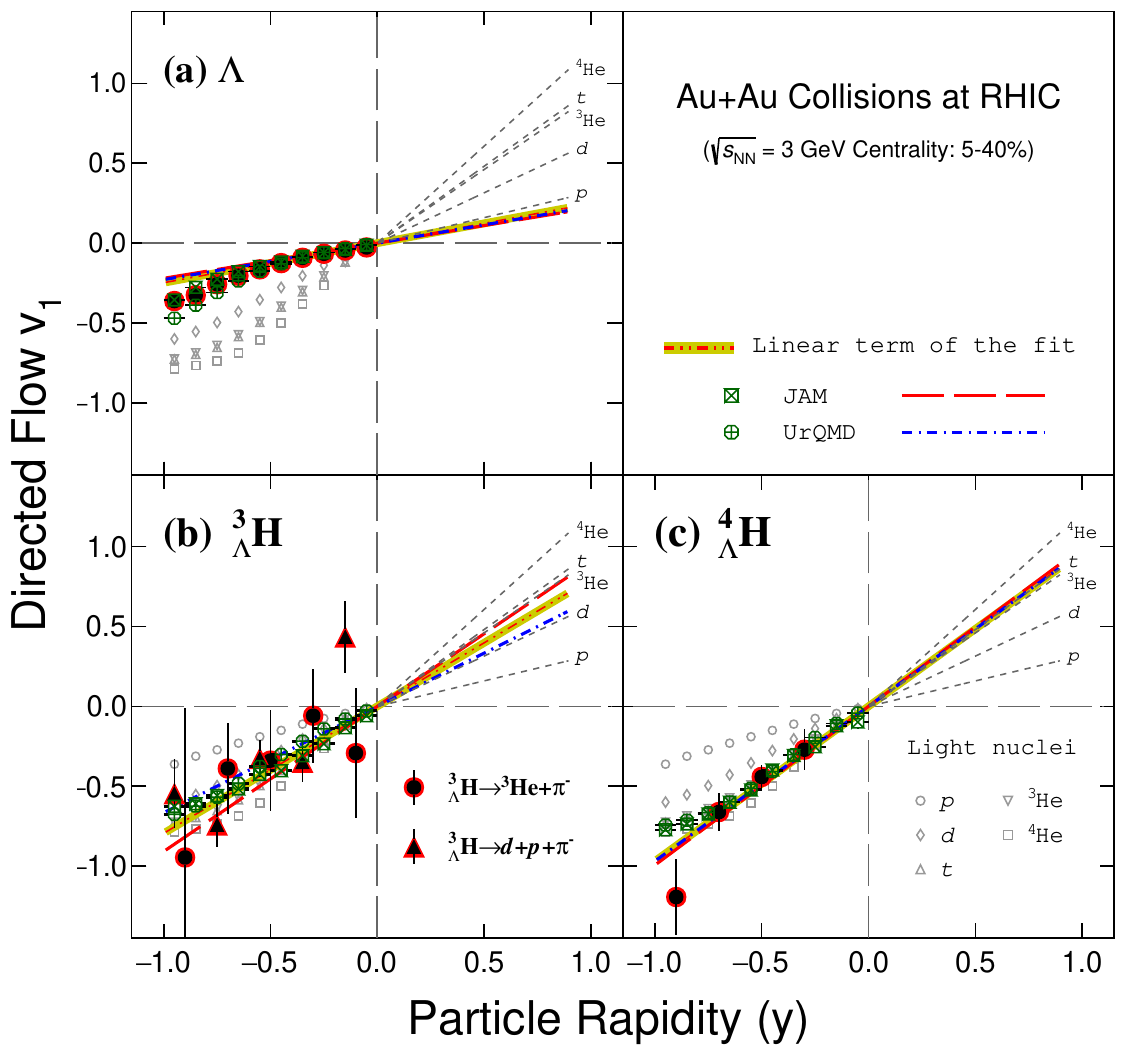}
\caption{$\Lambda$ hyperon and hypernuclei directed flow $v_1$, shown as a function of rapidity, from the \snn3 5-40\% mid-central Au + Au collisions. In case of $\HS$ $v_1$, both results from 2-body (circles) and 3-body (triangles) decays are shown. The fitted linear terms for light nuclei are plotted as dashed lines in the positive rapidity region, while for $\Lambda$ hyperon, $\HS$ and $\HF$, they are shown by the yellow-red lines in the corresponding panels. The rapidity dependence of $v_1$ for $p$, $d$, $t$, $^3$He, and $^4$He are also shown as open markers, including circles, diamonds, up-triangles, down-triangles and squares, respectively. The corresponding v$_1$ results extracted from 1st+3rd order polynomial fits, within $-1.0<y<0$, are shown as dashed lines in the positive rapidity region. Transport model calculations (fit results) from JAM and UrQMD are shown as cross circles and cross squares (long dashed and dot-dashed lines), respectively. }
\label{fig:model}
\end{figure}

Figure~\ref{fig:model} shows the $\Lambda$ hyperon and hypernuclei directed flow $v_1$, as a function of rapidity, from the \snn3 5-40\% mid-central Au + Au collisions.
The results of a linear fit: 
\begin{linenomath*}
\begin{equation}
v_{1}(y) = v_{1}^{s}{\cdot}{y} 
\end{equation}
\end{linenomath*}
for $\HS$ and $\HF$ are shown as red-yellow lines in the figure. The range of ($-1.0<y<0$) is used for v$_1$ measurements of all light nuclei, $\Lambda$ hyperon and hypernuclei, as well as their model results. As the $\rm \Lambda$ and light nuclei have an obvious non-linear tendency at $y<-0.5$. So a 1st plus 3rd order polynomial function
\begin{linenomath*}
\begin{equation}
v_{1}(y) = v_{1}^{s}{\cdot}{y}+p_1{\cdot}{y}^3
\end{equation}
\end{linenomath*}
is used to fit them. The same fitting procedure is applied to model calculations, as well as to the $\rm ^{3}_{\Lambda}H$ and $\rm ^{4}_{\Lambda}H$ model calculations. 
While for the $v_1$ results from $\rm ^{3}_{\Lambda}H$ and $\rm ^{4}_{\Lambda}H$, due to the limitation of their statistics, we use the 1st order polynomial function to describe them. In Fig. 5, transport model calculations (fit results) from JAM and UrQMD are shown as cross circles and cross squares (long dashed and dot-dashed lines), respectively.
The resulting mid-rapidity $v_{1}$ slopes for all the particles under study are summarized in Fig. 4 of the manuscript.

\clearpage
\bibliographystyle{unsrt}
\bibliography{supplemental}